# Machine learning unveils composition-property relationships in chalcogenide glasses

Saulo Martiello Mastelini[a], Daniel R. Cassar[b,1], Edesio Alcobaça[a], Tiago Botari[a], André C. P. L. F. de Carvalho[a], Edgar D. Zanotto[b]

[a]Institute of Mathematics and Computer Sciences, University of São Paulo, São Carlos, Brazil

[b]Department of Materials Engineering, Federal University of São Carlos, São Carlos, Brazil

## ABSTRACT

Due to their unique optical and electronic functionalities, chalcogenide glasses are materials of choice for numerous microelectronic and photonic devices. However, to extend the range of compositions and applications, profound knowledge about *composition-property* relationships is necessary. To this end, we collected a large quantity of composition-property data on chalcogenide glasses from SciGlass database regarding glass transition temperature ($T_g$), Young´s modulus ($E$), coefficient of thermal expansion (CTE), and refractive index ($n_D$). With these data, we induced predictive models using three machine learning algorithms: Random Forest, K-nearest Neighbors, and Classification and Regression Trees. Finally, the induced models were interpreted by computing the SHAP (SHapley Additive exPlanations) values of the chemical features, which revealed the key elements that significantly impacted the tested properties and quantified their impact. For instance, Ge and Ga increase $T_g$ and $E$ and decrease CTE (three properties that depend on bond strength), whereas Se has the opposite effect. Te, As, Tl, and Sb increase $n_D$ (which strongly depends on polarizability), whereas S, Ge, and P diminish it. Knowledge about the effect of each element on the glass properties is precious for semi-empirical compositional development trials or simulation-driven formulations. The induced models can be used to design novel chalcogenide glasses with required combinations of properties.

---

1    Corresponding author: contact@danielcassar.com.br

# Introduction

Chalcogenide glasses contain one or more chalcogens (sulfur, selenium, and tellurium) and no oxygen. Their lower band gaps ($E_g$ = 1–3 eV) lead to optical and electrical properties very different from those of oxide glasses ($E_g$ = 2.5–5 eV). This feature allows several high-technology applications that are not possible with other glass types. The unique functionalities of chalcogenide glasses make them the selected materials for microelectronic and photonic devices. They can be made as thin and thick films, molded into lenses, or drawn into fibers. They have been used in commercial applications, such as infrared cameras, fibers, laser waveguides for optical switching, and chemical and temperature sensors [1].

Chalcogenide compounds such as AgInSbTe and GeSbTe are also importantly applied in rewritable optical disks and phase-change memory devices. They are fragile glass formers according to Angell's classification [2]; by controlling heating and annealing (cooling), they can be very rapidly switched between non-crystalline and crystalline states, thereby significantly changing their optical and electrical properties and allowing information storage [1].

Chalcogenide glasses are traditionally composed of at least one chalcogen (Se, Te, and S) combined with Ge, As, Sb, Si, P, B, Pb, La, Al, or other neighboring atoms on the periodic table. Two characteristics of chalcogens provide chalcogenide glasses with unique properties: first, they generate low energy phonons within the non-crystalline network and confer wide optical transparency to glasses, extending far into the infrared (this property is a defining characteristic and has been the source for much research on infrared optics applications); second, several of these elements present similar electronegativity, close to 2, and, consequently, form directional covalent bonds.

Chalcogenide glasses are glassy semiconductors. There is relatively firm knowledge about their short-range structure, which covers the coordination number, the bond length, and the bond angle. Also, structural dependence on atomic compositions, which are practically possible in covalent glasses, has added valuable insights into the chalcogenide glass science [1].

The classical chalcogenide glasses (mainly sulfur-based, such as As-S or Ge-S) are reasonable glass-formers; however, their glass-forming abilities significantly decrease with increasing the molar weight of their constituent elements, i.e., S > Se > Te. Most of the formulations available are far worse glass-formers than the oxide compositions, and this is a critical issue in this glass family [1]. More recently, the glass research community started digging deeper into the crystallization behavior [3] and development of chalcogenide glass ceramics, keeping their optoelectronic properties and showing improved mechanical behavior [4].

A Scopus search (May 30, 2021) showed that approximately 10,000 articles addressing chalcogenide glasses have been published since the pioneering article by Kolomiets and Pishlo (1963) [5]; the current rate is about 400 articles per year. Due to incomplete structural knowledge (especially about medium-range structures, density fluctuations, and defects), the chalcogenide glass science is far behind those constructed for single-crystalline semiconductors or oxide glasses. Also, despite the substantial research conducted in the past 50 years, the understanding of composition-property relationships for chalcogenide glasses is still behind the accumulated knowledge about oxide glasses, which have been systematically studied by many researchers for approximately two centuries. Therefore, to extend the range of available compositions and applications of chalcogenide glasses, there is pressing need for more profound knowledge about the *composition-structure-property* relationships.

While the number of machine learning (ML) papers addressing oxide glasses has upsurged in the past five years, to the best of our knowledge, there is only one publication on ML research in chalcogenide glasses [6]. This study reports on a multivariate linear regression (MLR) capable of predicting the glass transition temperature of the $As_xSe_{1-x}$ binary system. The obtained MLR model ($T_g$ = 2464 + 597.3$\langle r \rangle$ − 6755.3$\nu$ − 301.61$K$ + 4.9257$U_{0ex}$ + 0.50313$KU_{0ex}$) agreed with experimental values for this particular binary system, and was based on physical and chemical properties such as the average coordination number $\langle r \rangle$, the Poisson ratio $\nu$, the bulk modulus $K$, and the mean experimental atomic bonding energy $U_{0ex}$.

The incentive of researching ML algorithms applied to chalcogenides was pointed out as an opportunity in the field by Tandia et al. [7]. Meeting this incentive is the main objective of this work. Here we use a completely different approach from that of Ref. [6]. We aim to *induce* ML models referring to *composition-property* relationships and interpret them to find the effect of each element on the glass properties. In addition, we will deal with much more complex compositions, containing up to six elements rather than with a single binary system. To this end, we collected published data regarding some critical properties of chalcogenide glasses: glass transition temperature ($T_g$), Young´s modulus ($E$), coefficient of thermal expansion (CTE), and refractive index ($n_D$), and use ML-based approaches to *generate predictive models* for these properties.

The following ML algorithms were tested in this study: CART (Classification And Regression Trees), k-NN (K-Nearest Neighbors) and RF (Random Forest), which were chosen because our previous work on oxide glasses indicated that these are the top performers among six ML algorithms [8]. We will check which of these algorithms performs the best. Finally, we will attempt to *interpret* the induced models of the RF by computing the SHAP (SHapley Additive exPlanations) values of the features, shedding light on the role that the chemical elements play in each property. We expect that the results of this work will help designing new chalcogenide glasses with desired combinations of properties.

# Methodology

## Data collection

The data on chalcogenide glasses used in this work were collected from the SciGlass database (https://github.com/epam). For a glass to be considered a chalcogenide for the purposes of the current simulation work, it had to meet two conditions. The first is having a non-zero amount of sulfur, selenium, or tellurium; the second is not having any amount of oxygen, nitrogen, fluorine, chlorine, bromine, iodine, gold, silver, platinum, and palladium. After this filtering procedure, only entries having one or more properties within the scope of this work (glass transition temperature, Young´s modulus, coefficient of thermal expansion, or refractive index) were considered. We also considered studying the Abbe number of chalcogenide glasses; however, only approximately 50 examples were available, which would not be enough for the proper use of ML algorithms.

The cleaning stage was performed to eliminate extremely low or high property values that likely refer to typos or gross measurement errors. The strategy used was similar to that employed in previous publications [9]: we removed the extreme values for each property and the duplicate entries by taking the median value of the property. All values below the 0.05% percentile or above the 99.95% percentile were defined as extreme. Descriptive statistics on the collected dataset are shown further in the text, in Table 1.

## Machine learning experiments

This work follows the same ML-based strategy we employed in a recent report on oxide glasses [8]. We considered three ML algorithms that performed well in a previous analysis, namely, CART, K-NN, and RF. These are well-known supervised ML algorithms. As such, detailed explanations on how they induce predictive models can be found elsewhere (see, for example, the supplementary material of Ref. [8]).

The predictive models were induced using the scikit-learn Python package [10]; a hyperparameter tuning routine was also employed. We adopted a nested cross-validation routine considering an outer-fold of 10 for testing and an inner-fold of 5 for validation. The tuning strategy was the use of random search, testing 500 sets of hyperparameters for each outer fold. Moreover, we used the same search space adopted in Ref. [9]. For experimental reproducibility, we make available the code used on GitHub (https://github.com/ealcobaca/mlglass).

## Interpreting the induced models through SHAP analysis

Models induced by ML algorithms can hold a significant amount of information, which may or may not be easily interpreted by humans (mainly depending on the used algorithm). A new and powerful data analysis tool called SHAP [11], distributed as a Python module (https://github.com/slundberg/shap), is a model-agnostic approach to interpret any predictive function and extract/visualize meaningful information in a human-readable fashion. The approach used by SHAP is the computation of the Shapley values [12], which are based on game theory and inform how much a given prediction is affected by the input features concerning a given base value. Detailed information on this procedure is reported by the creators of this method [11].

One possible way to visualize the results of the SHAP analysis is via beeswarm plots. These plots can be thought of as horizontal violin plots, with features sorted by decreasing order of importance. In this case, importance is measured by the absolute sum of the SHAP values, which indicates the features that have a higher impact on the predicted value of the model. The SHAP values have the same units of the property being predicted and convey how much a given feature (chemical elements in this case) impacts the property in relation to a base value, which is taken as the mean value of the property (see Table 1).

# Results and discussion

## Analysis of the datasets used in this study

Table 1 shows the descriptive statistics of the glass compositions collected from the SciGlass database. The smallest dataset was labeled with the refractive index (with 445 unique compositions), whereas the largest dataset was labeled with the glass transition temperature (with 6,747 unique compositions). While these numbers are much smaller than those used in the previous ML works on oxide glasses [9,13–18], they are still significant and can be used by ML algorithms to extract composition-property relations. It is relevant to note that current chalcogenide formulations comprise 51 elements, with only 1 to 6 in each glass.

**Table 1**. Descriptive statistics of the used datasets for each property.

|  | $T_g$ (K) | $E$ (GPa) | $\log_{10}$(CTE) | $n_D$ |
|---|---|---|---|---|
| **Count** | 6,747 | 479 | 865 | 445 |
| **Mean** | 479 | 19.1 | −4.68 | 2.61 |
| **Std Dev** | 111 | 6.0 | 0.21 | 0.41 |
| **Min** | 266 | 6.6 | −5.17 | 1.97 |
| **50%** | 457 | 18.4 | −4.72 | 2.50 |
| **Max** | 877 | 70.0 | −4.02 | 4.34 |
| **Skewness** | 0.73 | 2.82 | 0.57 | 1.04 |
| **Kurtosis** | 0.21 | 19.83 | −0.14 | 0.87 |

Figure 1 shows the histogram of the number of chemical elements in the glasses for each property, which varies from 1 to 6. These relatively "simple" compositions contrast with those of the widely studied oxide glasses, for which multi-component glasses with more than 20 elements are reported. Hopefully, this work could guide researchers in formulating novel multi-component chalcogenide glasses, as discussed further in this communication. Similarly, Fig. 2 shows the histogram for the property values, for which the minimum and maximum values can be found in Table 1. All studied properties have an asymmetric distribution, which is evidenced by the non-zero value for their skewness (Table 1).

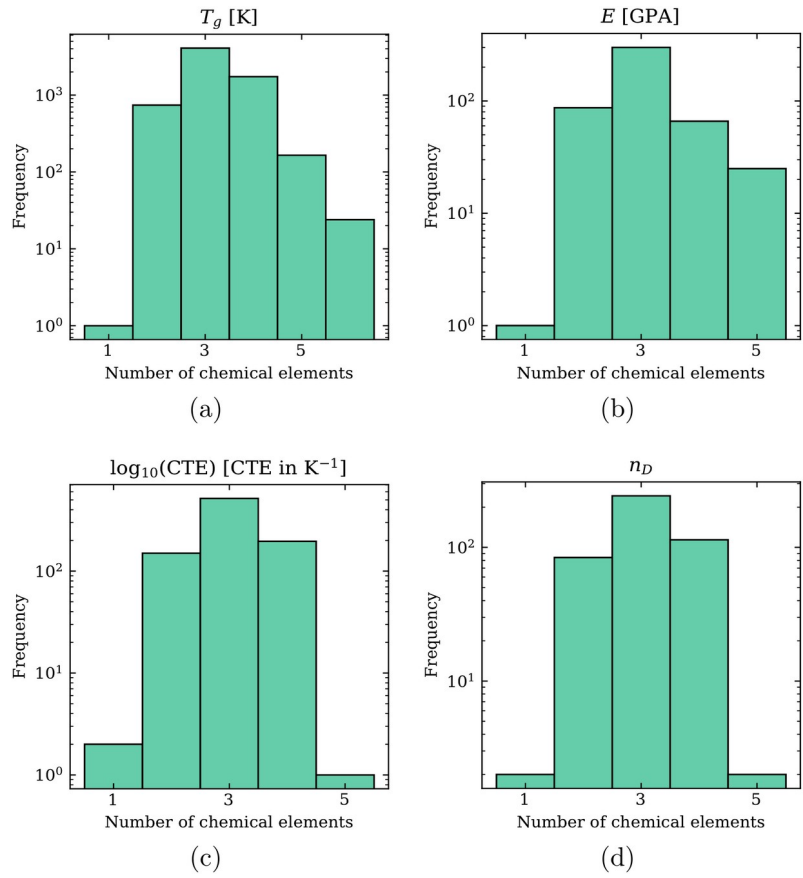

**Figure 1**. Frequency versus the number of chemical elements in each composition for four properties of chalcogenide glasses.

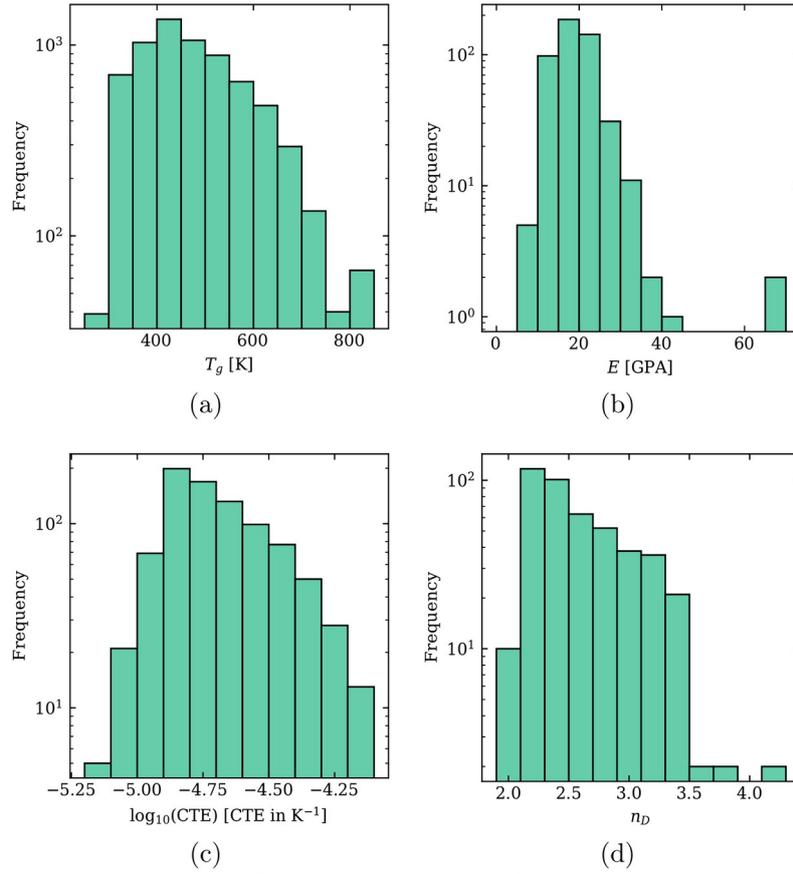

**Figure 2**. Frequency versus value for four properties of chalcogenide glasses.

## Predictive performance measures

Table 2 shows the predictive performance measures for the four properties obtained by the RF algorithm. Additional tables, with CART and k-NN results, can be found in the Appendix. In general, the predictive performance values obtained by RF and k-NN outperformed those obtained by CART. However, these metrics are not as good as those for the oxide glasses focused in our previous study [9], for which the number of examples (composition-property points) used in the training procedures was much larger (20,000–50,000). As expected, the uncertainty decreased with the number of examples used in the training procedure; for instance, $R^2$ is 0.86 for $n_D$ (445 examples) versus 0.93 for $T_g$ (6,747 examples).

**Table 2**. Values of the performance metrics for the four properties obtained using the tuned RF algorithm. The up arrow indicates that the higher the metric, the better; the down arrow indicates the opposite.

| Metric | $T_g$ (K) | $E$ (GPa) | $\log_{10}$(CTE) | $n_D$ |
|---|---|---|---|---|
| RD (↓) | $3.5 \pm 0.2$ | $9 \pm 1$ | $1.3 \pm 0.2$ | $3.3 \pm 0.7$ |
| R2 (↑) | $0.93 \pm 0.02$ | $0.66 \pm 0.18$ | $0.76 \pm 0.08$ | $0.86 \pm 0.05$ |
| RMSE (↓) | $29 \pm 3$ | $3.5 \pm 2.0$ | $0.10 \pm 0.02$ | $0.15 \pm 0.04$ |
| RRMSE (↓) | $0.26 \pm 0.03$ | $0.58 \pm 0.14$ | $0.49 \pm 0.08$ | $0.37 \pm 0.07$ |

Figure 3 shows the main results of the relative deviation of the $T_g$ prediction for the three ML algorithms used in the experiments. Again, as reported in previous communications [8,16,19], the uncertainty in the extremes of low and high $T_g$ is higher than in the intermediate range. This behavior is similar to those from other studied properties and reflects the small number of examples in the extreme regions. The plots for the other properties are reported in the Appendix.

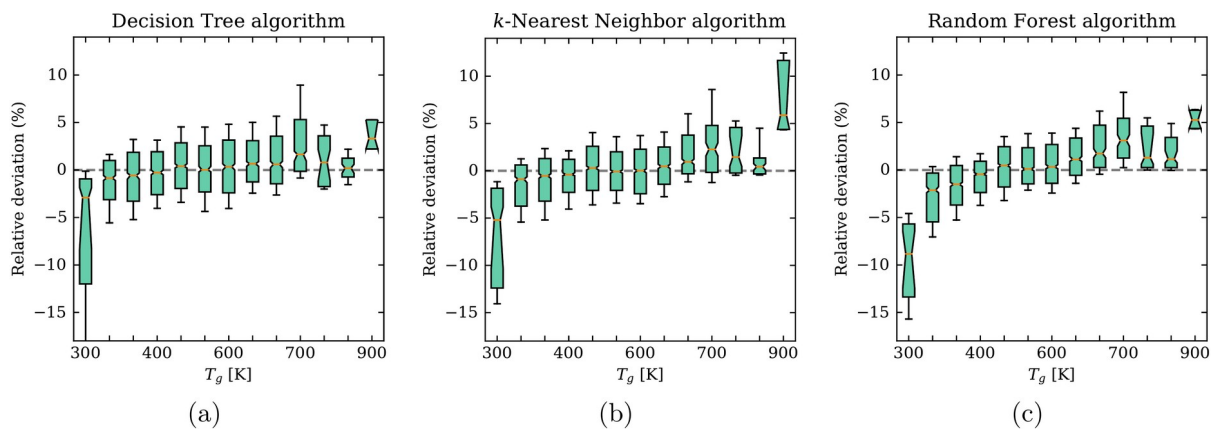

(a)            (b)            (c)

**Figure 3**. Boxplot of residuals for the prediction of $T_g$ for the tuned models. The boxes are bounded by the first and third quartiles, while the error bars comprehend 66% percent of the data. The mean is shown by a horizontal orange line and the notch represents its confidence interval.

Figure 4 shows the mean and standard deviation of the residual prediction (reported minus predicted values) of the $T_g$ model induced by RF for each chemical element in the glass. Again, this result is similar to those previously reported for oxide glasses [9] and confirms the expected behavior that the quality of prediction improves as the number of examples increases. Elements that are part of a larger number of glass compositions tend to have a mean residual prediction close to zero.

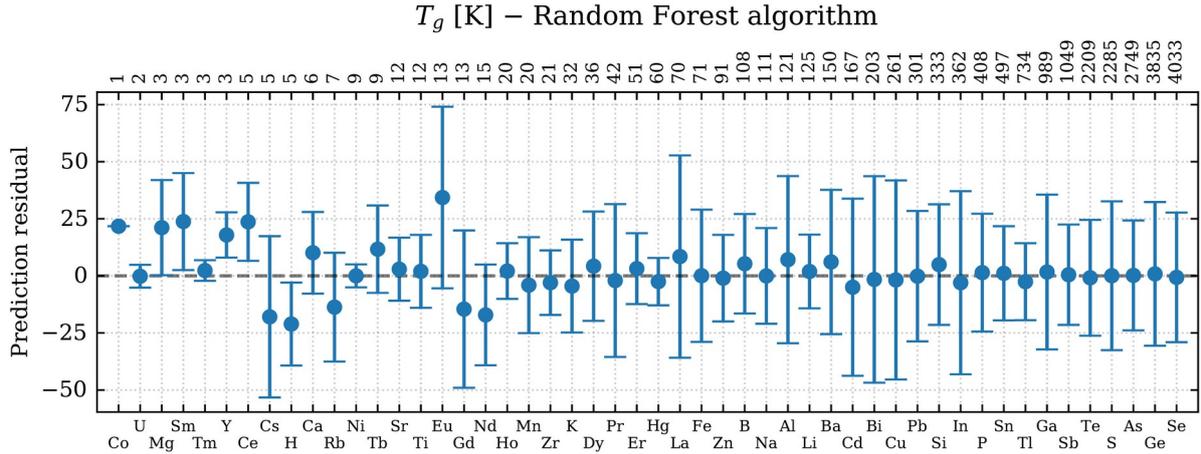

**Figure 4**. Mean and standard deviation of the prediction residual of $T_g$ for each chemical element. The numbers in the top are the number of examples (glass compositions) containing that element in the dataset.

The induced predictive models can be used for the computer-aided design of new chalcogenide glasses having desired combinations of properties. However, due to the limited dataset used for training these models, unsatisfactory predictions will likely result on searching for chemical compositions that contain certain elements that are present in small number of compositions, such as Co, U, Mg, Sm, Tm, Y, Ce, Cs, H and a few others shown in Fig. 4. The same restriction applies for new formulations that are far away from those present in the training dataset. To mitigate this problem, we would have to significantly extend the dataset.

In the following section we will dig deeper into the RF induced models in an attempt to extract useful information regarding the effect of each chemical element on the properties. To this end, we will use the SHAP analysis discussed in the methodology section.

## Interpreting the induced models

By employing the SHAP analysis, we obtained the plots shown in Figure 5 for the four studied properties. Although the SHAP still presents some problems [20,21], these figures provide valuable insights for designing chalcogenide glasses. Each dot in these plots represents a glass having the chemical element shown in the respective left label (note that the dots can stack vertically, conveying the message that many glasses have the same SHAP value). The x-axis shows the SHAP values, which have the same units of the target property and quantify the impact of the feature (chemical element) on the property. Finally, each dot has a color representing the atomic fraction of the element in the glass.

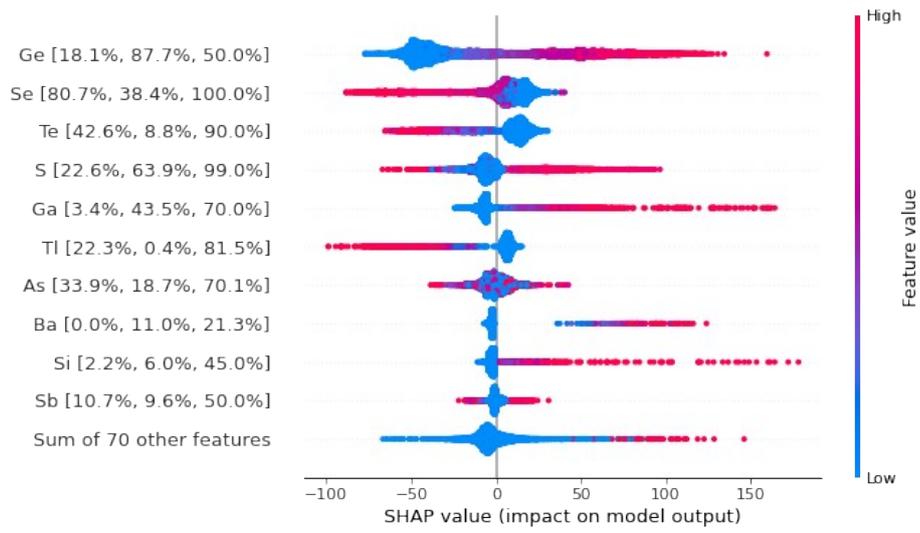

**(a)**

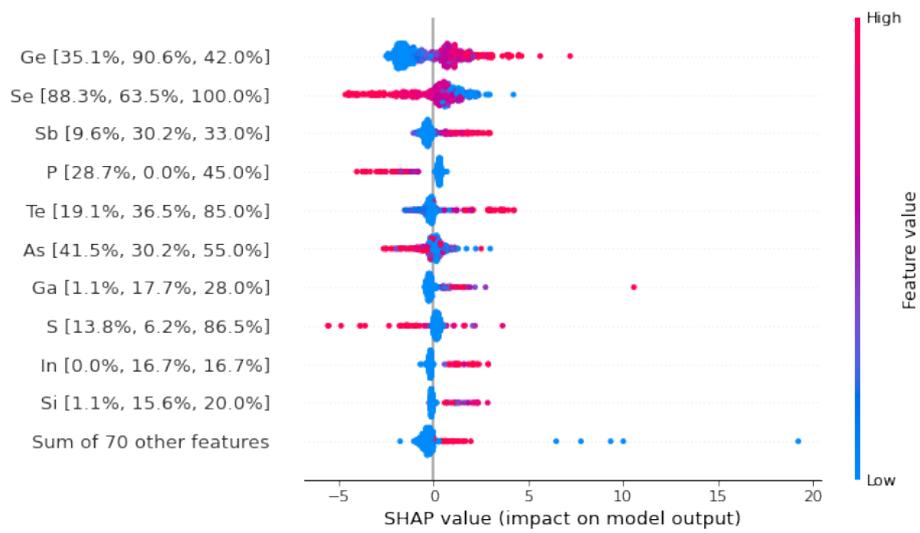

**(b)**

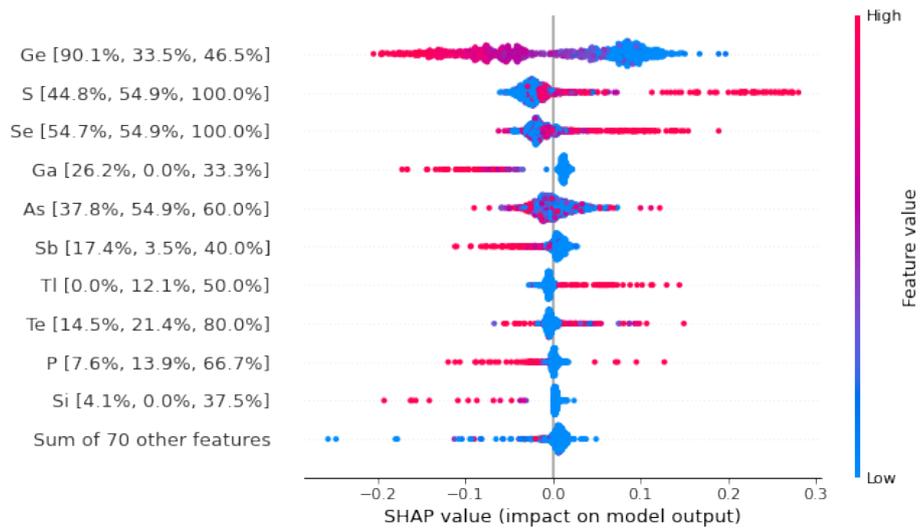

**(c)**

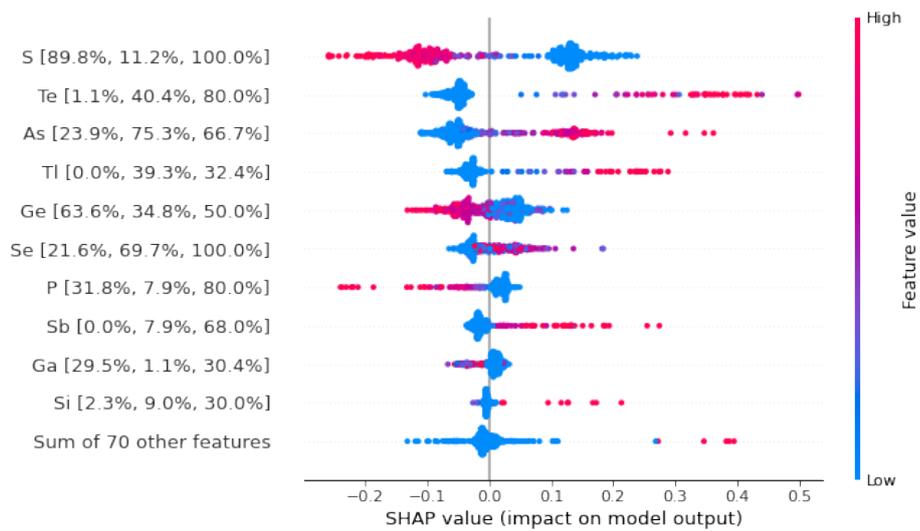

**(d)**

**Figure 5**: Beeswarm plot of the SHAP values obtained from the RF predictive model of (**a**) $T_g$, (**b**) $E$, (**c**) $\log_{10}(\text{CTE})$, and (**d**) $n_D$. The numbers within brackets beside the chemical element labels represent, respectively, the percentage of examples that contain the said element in the low range of the property (lower than the 20% percentile), the percentage of examples that contain the said element in the high range of the property (higher than the 80% percentile), and the maximum atomic fraction of the element in the dataset.

Figure 5a shows that large amounts of germanium (Ge), gallium (Ga), barium (Ba), and silicon (Si) contribute to increasing the $T_g$ of chalcogenide glasses, whereas selenium (Se), tellurium (Te), and thallium (Tl) contribute to decreasing it. Sulfur (S), arsenic (As), and antimony (Sb) have a mixed effect; they can increase or decrease this property, suggesting that these elements interact with the glass network in a complex way.

Figure 5b shows that high percentages of germanium, antimony, tellurium, gallium, indium (In), and silicon contribute to increasing $E$, whereas selenium, arsenic, and phosphorus (P) contribute to decreasing it. Sulfur shows a mixed effect.

Finally, Figure 5c shows that sulfur, selenium, and thallium increase CTE, while germanium, gallium, antimony, and silicon contribute to decreasing it. A mixed effect is observed for arsenic, tellurium, and phosphorus for this property.

The three properties discussed in the previous two paragraphs ($T_g$, $E$, and CTE) are related to the chemical bond energy. They are highly affected by germanium and selenium, elements that rank high in the SHAP importance analysis.

The refractive index analysis, shown in Figure 5d, reveals a different picture, which is expected as this property is not directly related to the chemical bond energies, but to the polarizability of the elements. Here, sulfur is the most important, and it can either increase $n_D$ when present in small quantities or decrease it, when present in large quantities. Tellurium, arsenic, thallium, selenium, antimony, and silicon increase the refractive index in chalcogenide glasses, while germanium, phosphorus, and gallium decrease it. No mixed effect was observed in the features shown in Figure 5d.

Now, we will look into more detail on the magnitude of the SHAP values, which, as already mentioned, quantifies the impact of the elements on the final prediction of the model. Starting by Fig. 5a, we see that germanium (87.7%), gadolinium (43.5%), and silicon (6%) can rise $T_g$ the most, up to about 170 K. The numbers in parentheses refer to the percentage of high $T_g$ glasses (above the 80% percentile) in the dataset containing these chemicals. As it can be seen, by simply looking at the reported chalcogenide glasses having high $T_g$, one can miss the significant impact of silicon on this property.

Similarly, selenium (80.7%), tellurium (42.6%), and thallium (22.3%) are elements that can decrease $T_g$ the most, down by 100 K in the most extreme case. The numbers in parentheses refer to the percentage of low $T_g$ glasses (below the 20% percentile) in the dataset containing these chemicals. As previously mentioned, these analyses provide us with rich information to empirically design new chalcogenide glasses. The following paragraphs explore the other three properties, with the numbers in parentheses having the same meaning.

Figure 5b shows that germanium (90.6%), tellurium (36.5%), and antimony (30.2%) can increase $E$ the most, an increase of approximately 9 GPa in the most extreme case, but usually staying below 5 GPa. Selenium (88.3%), phosphorus (28.7%), and arsenic (41.5%) can decrease $E$ by up to 5 GPa. As expected, this analysis agrees with Fig. DATA_3c, which shows that there is no significant variance in $E$ among the known chalcogenide glasses.

Figure 5c shows that sulfur (54.9%), selenium (54.9%), and germanium (33.5%) are the elements with the most significant impact on increasing CTE, which can amount to 0.3 in base-10 logarithm scale for the most extreme case. Interestingly, germanium only increases CTE when present in small quantities, but even so, it has a significant impact on this property. Germanium (90.1%), gallium (26.2%), and silicon (4.1%) play the most significant role in decreasing CTE, reaching up to 0.2 in the base-10 logarithm scale.

Finally, Figure 5d shows that tellurium (40.4%), arsenic (75.3%), and thallium (39.3%) can significantly increase $n_D$, the first reaching an impressive impact of 0.5 on this property. Meanwhile, sulfur (89.8%) and phosphorus (31.8%) can decrease this property by more than 0.2.

The above discussion shows that analyses of SHAP plots can reveal the effect of the chemical elements on the properties and the respective magnitudes.

# Summary and conclusions

In this study, we collected over six thousand composition-property sets for four properties of chalcogenide glasses. Current chalcogenide formulations comprise 51 chemical elements, with 1 to 6 elements in each glass. We used these data to train and test three different ML algorithms, for this important glass family for the first time. The RF and k-NN algorithms outperformed the CART algorithm in predictive performance, confirming previous results for oxide glasses.

A SHAP analysis of the RF models indicated the key elements that significantly increase or decrease the value of the tested properties and their maximum possible variation. For instance: germanium increases $T_g$ and $E$ and decreases CTE. This occurs likely because these elements increase the interatomic bond strength of these covalent glasses, whereas selenium has the opposite effect on these properties. Tellurium, arsenic, thallium, and antimony increase $n_D$, which depends mostly on polarizability, whereas sulfur and phosphorus diminish it.

This knowledge about the effect of each element on properties can be precious for semi-empirical compositional development trials of chalcogenide glasses. Besides, the induced predictive models can be used for the computer-aided design of new chalcogenide glasses having desired combinations of properties. However, due to the limited dataset used for training these models, unsatisfactory predictions will likely result on searching for chemical compositions that are too far away from those present in the training dataset. The same restriction applies to other substances, such as oxide, metallic, and organic glasses. To mitigate this problem, the research community will have to significantly extend the composition-property dataset.

**Acknowledgments**

The São Paulo Research Foundation (FAPESP) financed this study through grants nos. 2017/12491-0, 2018/07319-6, 2017/06161-7, 2018/14819-5, 2013/07375-0, and 2013/07793-6.

**Competing interest statement**

The authors declare no competing financial or non-financial interests.

**CRediT author statement**

**Saulo Martiello Mastelini**: Methodology, Software, Investigation, Data Curation, Writing - Review & Editing. **Daniel R. Cassar**: Software, Formal Analysis, Data Curation, Writing - Original Draft, Writing - Review & Editing, Visualization. **Edesio Alcobaça**: Methodology, Software, Data Curation, Writing - Review & Editing, Visualization. **Tiago Botari**: Methodology, Software, Data Curation, Writing - Review & Editing. **André C.P.L.F. de Carvalho**: Resources, Data Curation, Writing - Review & Editing, Supervision, Funding acquisition. **Edgar D. Zanotto**: Conceptualization, Formal Analysis, Data Curation, Writing - Original Draft, Writing - Review & Editing, Supervision, Funding acquisition.

# Appendix

Tables A.1 and A.2 show the performance measures for the CART and K-NN algorithms. Tables A.3 to A.6 show the metrics for the induced models for the four properties studied in this work. Finally, Figs. A.1 to A.6 show the boxplots and the residual plots vs. chemical elements for $E$, $\log_{10}(CTE)$, and $n_D$.

**Table A.1**. Values of the performance metrics for the four properties obtained using the tuned CART algorithm. The up arrow indicates that the higher the metric, the better; the down arrow indicates the opposite.

| Metric | $T_g$ (K) | $E$ (GPa) | $\log_{10}(CTE)$ | $n_D$ |
|---|---|---|---|---|
| RD ($\downarrow$) | $4.6 \pm 0.3$ | $12 \pm 2$ | $1.71 \pm 0.26$ | $4.6 \pm 1.1$ |
| R2 ($\uparrow$) | $0.88 \pm 0.02$ | $0.52 \pm 0.21$ | $0.64 \pm 0.11$ | $0.77 \pm 0.10$ |
| RMSE ($\downarrow$) | $39 \pm 4$ | $4.2 \pm 2.1$ | $0.13 \pm 0.02$ | $0.21 \pm 0.07$ |
| RRMSE ($\downarrow$) | $0.35 \pm 0.04$ | $0.72 \pm 0.16$ | $0.62 \pm 0.11$ | $0.50 \pm 0.12$ |

**Table A.2**. Values of the performance metrics for the four properties obtained using the tuned k-NN algorithm. The up arrow indicates that the higher the metric, the better; the down arrow indicates the opposite.

| Metric | $T_g$ (K) | $E$ (GPa) | $\log_{10}(CTE)$ | $n_D$ |
|---|---|---|---|---|
| RD ($\downarrow$) | $3.8 \pm 0.2$ | $9.7 \pm 1.5$ | $1.3 \pm 0.2$ | $3.2 \pm 0.8$ |
| R2 ($\uparrow$) | $0.92 \pm 0.02$ | $0.57 \pm 0.22$ | $0.76 \pm 0.07$ | $0.87 \pm 0.05$ |
| RMSE ($\downarrow$) | $32 \pm 3$ | $3.9 \pm 2.2$ | $0.10 \pm 0.02$ | $0.15 \pm 0.04$ |
| RRMSE ($\downarrow$) | $0.29 \pm 0.03$ | $0.65 \pm 0.15$ | $0.48 \pm 0.08$ | $0.36 \pm 0.08$ |

**Table A.3**: Experimental results for $T_g$.

| Metric | Cart | | k-NN | | RF | |
|--------|------|------|------|------|------|------|
| | Default | Tuning | Default | Tuning | Default | Tuning |
| RD | $4.6 \pm 0.3$ | $4.6 \pm 0.2$ | $4.0 \pm 0.2$ | $3.8 \pm 0.2$ | $3.6 \pm 0.2$ | $3.5 \pm 0.2$ |
| $R^2$ | $0.88 \pm 0.03$ | $0.88 \pm 0.02$ | $0.91 \pm 0.02$ | $0.92 \pm 0.02$ | $0.92 \pm 0.02$ | $0.93 \pm 0.02$ |
| RMSE | $39.6 \pm 3.9$ | $39.2 \pm 3.5$ | $32.57 \pm 3.06$ | $31.7 \pm 3.2$ | $31.0 \pm 3.0$ | $29.1 \pm 3.0$ |
| RRMSE | $0.36 \pm 0.04$ | $0.35 \pm 0.04$ | $0.29 \pm 0.03$ | $0.29 \pm 0.03$ | $0.28 \pm 0.03$ | $0.26 \pm 0.03$ |

**Table A.4**: Experimental results for $E$.

| Metric | Cart | | k-NN | | RF | |
|--------|------|------|------|------|------|------|
| | Default | Tuning | Default | Tuning | Default | Tuning |
| RD | $11 \pm 2$ | $12 \pm 2$ | $10 \pm 1$ | $10 \pm 2$ | $9 \pm 1$ | $9 \pm 1$ |
| $R^2$ | $0.5 \pm 0.2$ | $0.5 \pm 0.2$ | $0.6 \pm 0.2$ | $0.6 \pm 0.2$ | $0.6 \pm 0.2$ | $0.7 \pm 0.2$ |
| RMSE | $4 \pm 2$ | $4 \pm 2$ | $4 \pm 2$ | $4 \pm 2$ | $4 \pm 2$ | $3 \pm 2$ |
| RRMSE | $0.7 \pm 0.2$ | $0.7 \pm 0.2$ | $0.6 \pm 0.2$ | $0.7 \pm 0.2$ | $0.6 \pm 0.2$ | $0.6 \pm 0.1$ |

**Table A.5**: Experimental results for $\log_{10}(CTE)$.

| Metric | Cart | | k-NN | | RF | |
|--------|------|------|------|------|------|------|
| | Default | Tuning | Default | Tuning | Default | Tuning |
| RD | $1.6 \pm 0.2$ | $1.7 \pm 0.3$ | $1.3 \pm 0.2$ | $1.3 \pm 0.2$ | $1.3 \pm 0.2$ | $1.3 \pm 0.2$ |
| $R^2$ | $0.6 \pm 0.1$ | $0.6 \pm 0.1$ | $0.76 \pm 0.08$ | $0.76 \pm 0.07$ | $0.74 \pm 0.09$ | $0.76 \pm 0.08$ |
| RMSE | $0.13 \pm 0.02$ | $0.13 \pm 0.02$ | $0.10 \pm 0.02$ | $0.10 \pm 0.01$ | $0.10 \pm 0.02$ | $0.10 \pm 0.02$ |
| RRMSE | $0.6 \pm 0.1$ | $0.6 \pm 0.1$ | $0.49 \pm 0.09$ | $0.48 \pm 0.08$ | $0.51 \pm 0.09$ | $0.49 \pm 0.08$ |

**Table A.6**: Experimental results for $n_D$.

| Metric | Cart | | k-NN | | RF | |
|--------|------|------|------|------|------|------|
| | Default | Tuning | Default | Tuning | Default | Tuning |
| RD | $4.4 \pm 1.2$ | $4.6 \pm 1.1$ | $3.6 \pm 0.9$ | $3.2 \pm 0.8$ | $3.5 \pm 0.7$ | $3.3 \pm 0.7$ |
| $R^2$ | $0.77 \pm 0.09$ | $0.77 \pm 0.10$ | $0.85 \pm 0.06$ | $0.87 \pm 0.05$ | $0.85 \pm 0.06$ | $0.86 \pm 0.05$ |
| RMSE | $0.21 \pm 0.08$ | $0.21 \pm 0.07$ | $0.16 \pm 0.04$ | $0.15 \pm 0.04$ | $0.16 \pm 0.04$ | $0.15 \pm 0.04$ |
| RRMSE | $0.5 \pm 0.1$ | $0.5 \pm 0.1$ | $0.39 \pm 0.08$ | $0.36 \pm 0.08$ | $0.39 \pm 0.07$ | $0.37 \pm 0.07$ |

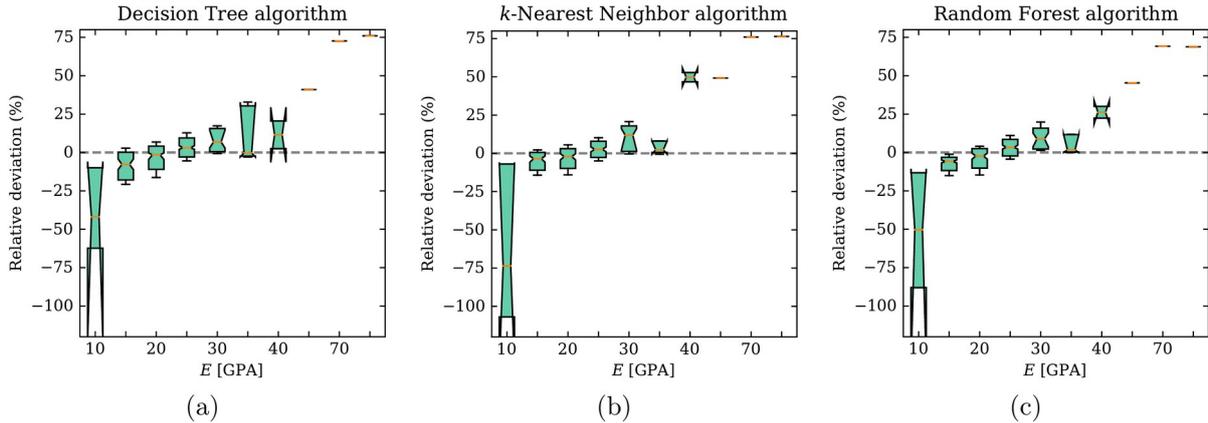

**Figure A.1.** Boxplot of residuals for the prediction of *E* for the tuned models. The boxes are bounded by the first and third quartiles, while the error bars comprehend 66% percent of the data. The mean is shown by a horizontal orange line and the notch represents its confidence interval.

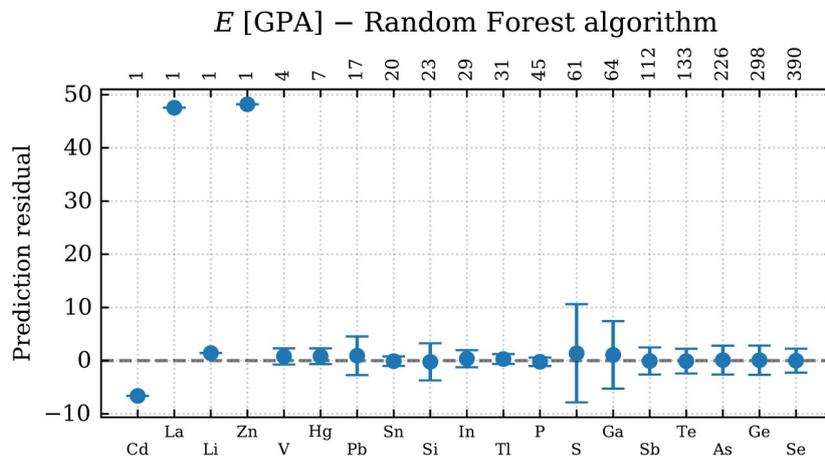

**Figure A.2.** Mean and standard deviation of the prediction residual of *E* for each chemical element. The numbers in the top are the number of examples (glass compositions) containing that element in the dataset.

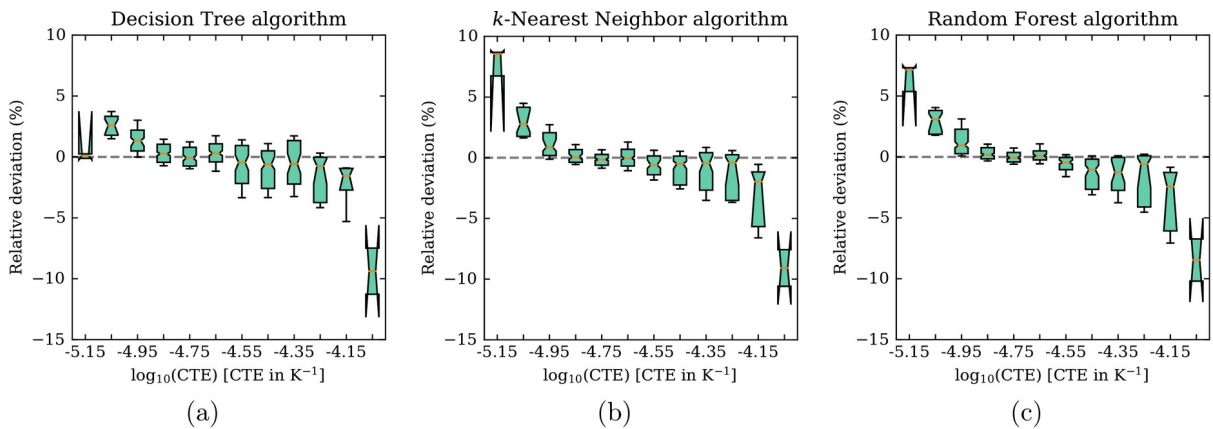

**Figure A.3.** Boxplot of residuals for the prediction of $\log_{10}$(CTE) for the tuned models. The boxes are bounded by the first and third quartiles, while the error bars comprehend 66% percent of the data. The mean is shown by a horizontal orange line and the notch represents its confidence interval.

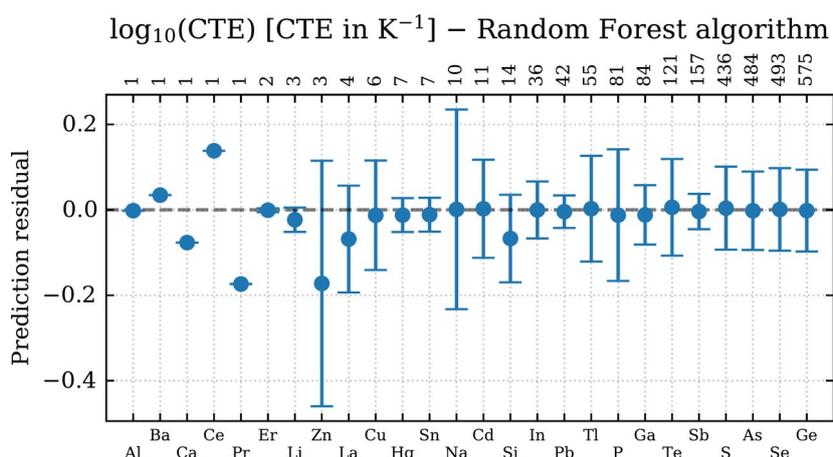

**Figure A.4.** Mean and standard deviation of the prediction residual of $\log_{10}(CTE)$ for each chemical element. The numbers in the top are the number of examples (glass compositions) containing that element in the dataset.

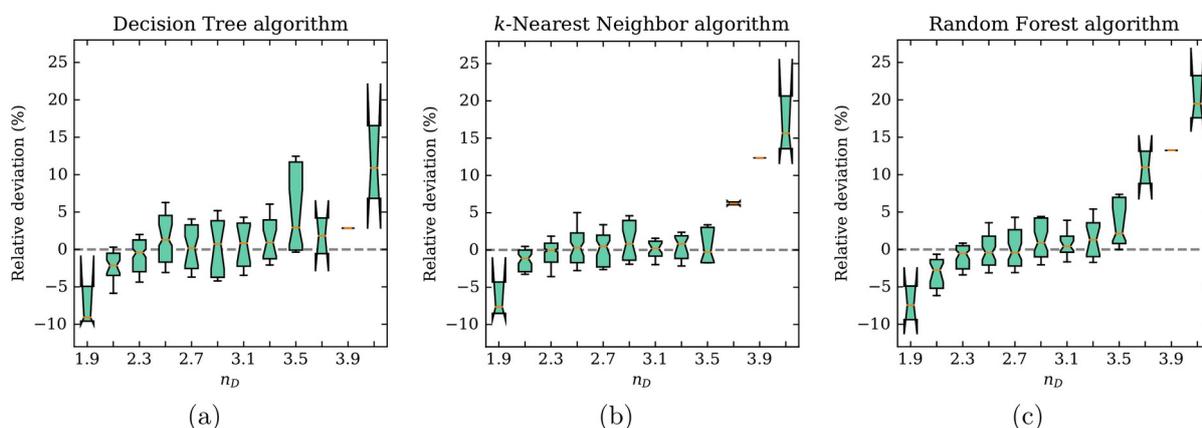

**Figure A.5**. Boxplot of residuals for the prediction of $n_D$ for the tuned models. The boxes are bounded by the first and third quartiles, while the error bars comprehend 66% percent of the data. The mean is shown by a horizontal orange line and the notch represents its confidence interval.

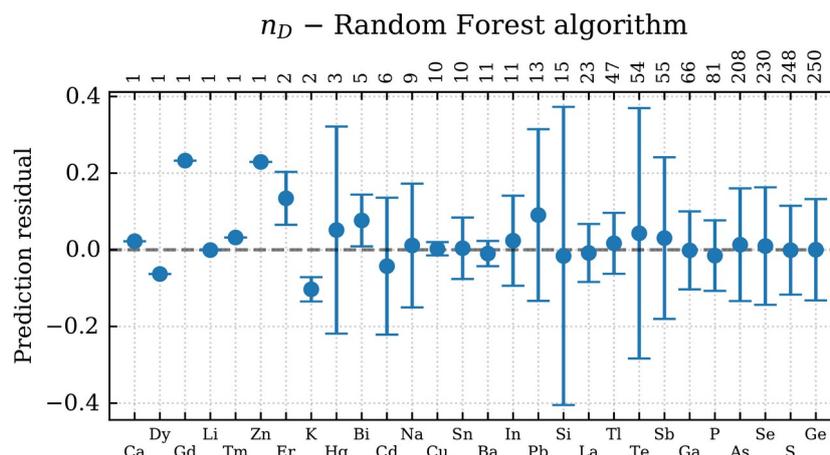

**Figure A.6**. Mean and standard deviation of the prediction residual of $n_D$ for each chemical element. The numbers in the top are the number of examples (glass compositions) containing that element in the dataset.